\documentclass[%
 aip,
 amsmath,amssymb,
 reprint,%
]{revtex4-1}

\usepackage{bm,latexsym,mathrsfs,enumerate}
\usepackage[table,dvipsnames]{xcolor}
\usepackage{upgreek}
\usepackage[breaklinks=true,unicode=true,urlcolor = blue,colorlinks = true,citecolor = blue,linkcolor = blue]{hyperref}
\usepackage{graphicx}
\usepackage{dcolumn}
\graphicspath{{./}}
\usepackage{mathtools}
\usepackage{todonotes}

\graphicspath{{./}}
\renewcommand{\vec}[1]{\bm{#1}}

\usepackage{mathptmx}

\usepackage[final]{pdfpages}
\makeatletter
\AtBeginDocument{\let\LS@rot\@undefined}
\makeatother

\begin{document}
\preprint{\textcolor[rgb]{0.00,0.50,0.75}{{\texttt{Draft \gitAbbrevHash{} by \gitCommitterName{} on \gitCommitterDate}}}}

\title{Unidirectional tilt of domain walls in equilibrium in biaxial stripes with Dzyaloshinskii--Moriya interaction}

\author{Oleksandr V. Pylypovskyi}
\email{o.pylypovskyi@hzdr.de}
\affiliation{Helmholtz-Zentrum Dresden-Rossendorf e.V., Institute of Ion Beam Physics and Materials Research, 01328 Dresden, Germany}
\affiliation{Taras Shevchenko National University of Kyiv, 01601 Kyiv, Ukraine}

\author{Volodymyr P. Kravchuk}
\email{volodymyr.kravchuk@kit.edu}
\affiliation{Institut f\"{u}r Theoretische Festk\"{o}rperphysik, Karlsruher Institut f\"{u}r Technologie, D-76131 Karlsruhe, Germany}
\affiliation{Bogolyubov Institute for Theoretical Physics of National Academy of Sciences of Ukraine, 03143 Kyiv, Ukraine}

\author{Oleksii M. Volkov}
\email{o.volkov@hzdr.de}
\affiliation{Helmholtz-Zentrum Dresden-Rossendorf e.V., Institute of Ion Beam Physics and Materials Research, 01328 Dresden, Germany}

\author{J\"{u}rgen Fa\ss bender}
\email{j.fassbender@hzdr.de}
\affiliation{Helmholtz-Zentrum Dresden-Rossendorf e.V., Institute of Ion Beam Physics and Materials Research, 01328 Dresden, Germany}

\author{Denis D. Sheka}
\email{sheka@knu.ua}
\affiliation{Taras Shevchenko National University of Kyiv, 01601 Kyiv, Ukraine}

\author{Denys Makarov}
\email{d.makarov@hzdr.de}
\affiliation{Helmholtz-Zentrum Dresden-Rossendorf e.V., Institute of Ion Beam Physics and Materials Research, 01328 Dresden, Germany}
\date{\textcolor[rgb]{0.00,0.50,0.75}{10 January 2020}}




\begin{abstract}
The orientation of a chiral magnetic domain wall in a racetrack determines its dynamical properties. In equilibrium, magnetic domain walls are expected to be oriented perpendicular to the stripe axis. We demonstrate the appearance of a unidirectional domain wall tilt in out-of-plane magnetized stripes with biaxial anisotropy and Dzyaloshinskii--Moriya interaction (DMI). The tilt is a result of the interplay between the in-plane easy-axis anisotropy and DMI. We show that the additional anisotropy and DMI prefer different domain wall structure: anisotropy links the magnetization azimuthal angle inside the domain wall with the anisotropy direction in contrast to DMI, which prefers the magnetization perpendicular to the domain wall plane. Their balance with the energy gain due to domain wall extension defines the equilibrium magnetization the domain wall tilting. We demonstrate that the Walker field and the corresponding Walker velocity of the domain wall can be enhanced in the system supporting tilted walls. 
\end{abstract}

\maketitle


Spin orbitronics relies on the manipulation of magnetic textures via spin orbit torques and enables new devices ideas for application in magnetic storage and logics\cite{Manchon14,Kuschel15,Chen17,Zhang16d,Divinskiy18}. The key component of these devices is a stripe with out of plane easy axis of magnetization and featuring the Dzyaloshinskii--Moriya interaction (DMI). Typically, asymmetrically sandwiched ultrathin films of ferromagnets are used, where the DMI originates from the broken symmetry at the film interfaces\cite{Fert80,Crepieux98}. There are numerous demonstrations of the energy efficient and fast motion of chiral magnetic solitons including skyrmions\cite{Fert17,Zhang18e,Goebel19a}, skyrmion-bubbles\cite{Jiang15,Woo16} and domain walls\cite{Parkin08,Ryu13,Emori13} in stripes. 

The orientation of the plane of a domain wall with respect to the stripe main axis has major impact on its dynamics including the maximum velocity\cite{Baumgartner18} and Walker limit\cite{Slastikov19,Vandermeulen16}. It is commonly accepted that in equilibrium the plane of a magnetic domain wall is perpendicular to the stripe main axis. This remains true even if the sample possesses DMI. Domain wall can acquire a tilt yet only if exposed to an external magnetic field\cite{Boulle13a,Emori14,Tetienne15,Muratov17,Shen19b}, driven by a current\cite{Viret05,Yamanouchi06,Ryu12,Boulle13a,Yu16,Baumgartner18}, or pinned on edge roughness during current-induced dynamics\cite{Martinez14}. Being exposed to an in-plane magnetic field, domain wall tilts unidirectionaly with the rotation direction determined by the sign of the DMI. The tilt increases linearly with the field strength and the slope of the resulting dependence was proposed to be used for the determination of the DMI constant\cite{Boulle13a,Muratov17}.

Here, we demonstrate that domain walls can acquire a unidirectional tilt even at equilibrium if the out-of-plane magnetized stripe possesses DMI and an additional easy-axis anisotropy in the plane of the stripe. The easy axis direction of the in-plane anisotropy can be given by a crystalline structure of the ferromagnet\cite{Heide08} 
or induced via exchange bias when the stripe is in proximity to an antiferromagnet\cite{Jimenez11,Camarero05,Kim19b,Zhang16d}. In contrast to the shape anisotropy\cite{Yan17,Slastikov19} with an easy axis along the stripe, any misalignment between the in-plane anisotropy and the stripe axes breaks the symmetry of the magnetic texture and tilts (i) the magnetization inside the domain wall as well as (ii) the plane of the domain wall. The motion a domain wall in a biaxial magnetic with DMI has strong impact on the Walker field and the domain wall velocity. The obtained results allow to design stripes with stronger Walker field (i.e. extend range of the linear motion of domain walls) and control the domain wall nucleation process in T-junctions by selecting the initial tilt direction\cite{Kwon17,Kwon18}.


\begin{figure*}
	\includegraphics[width=0.95\textwidth]{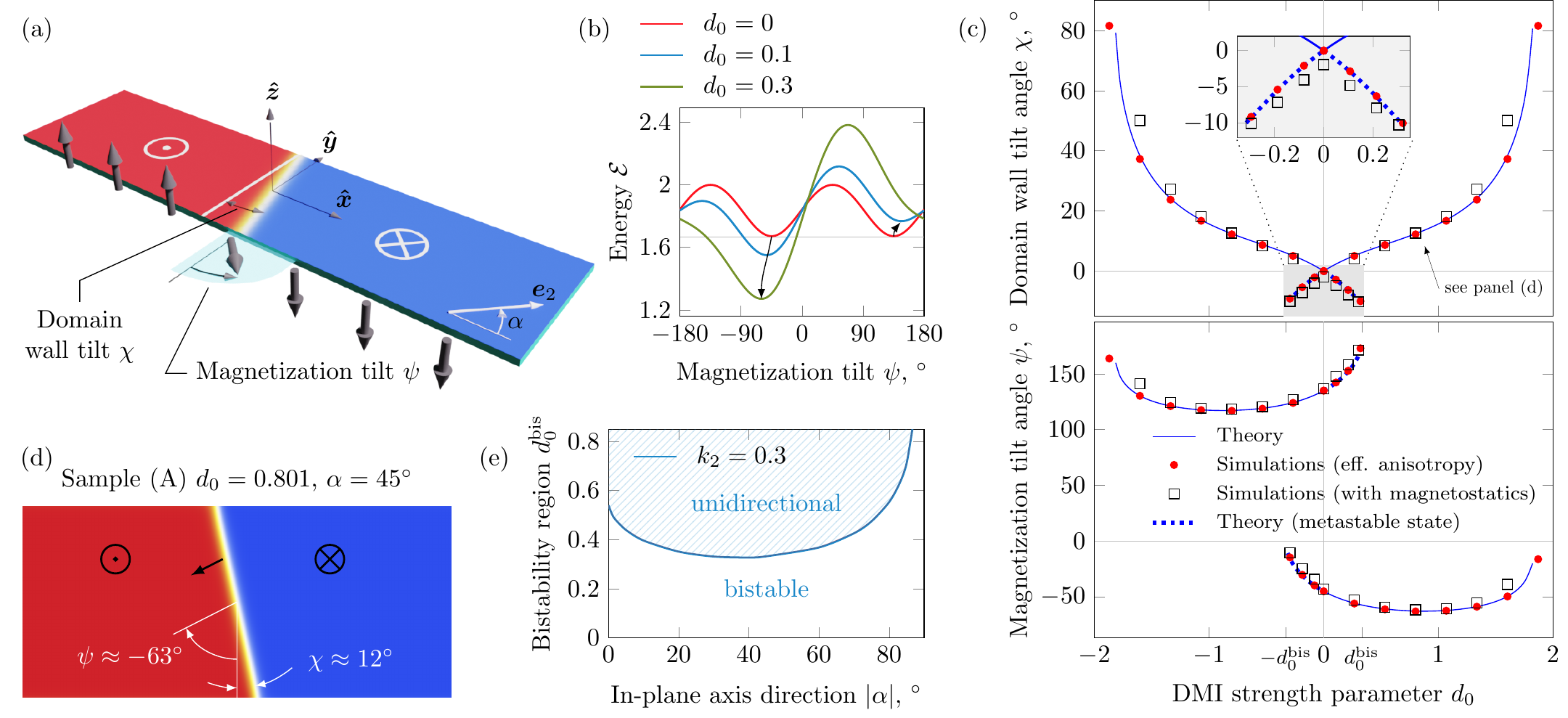}
	\vspace{-3mm}
	\caption{\textbf{The equilibrium structure and orientation of a domain wall in a biaxial stripe with DMI.} (a) Schematics of the out-of-plane magnetized stripe containing a domain wall. The mechanical tilt of the domain wall with respect to the $\vec{\hat{y}}$ axis is characterized with $\chi$. Angle $\psi$ describes the tilt of the magnetization in the wall. (b) Energy profile (Eq.~\eqref{eq:etot-psi}) for the case of $k_2 = 0.30$, $\alpha=45^\circ$, $p=+1$ and varying strength of the DMI, $d_0$. Arrows show the evolution of the energy minima with the change of DMI strength parameter $|d_0|$. 
	(c) Domain wall tilt angle ($\chi$) and magnetization tilt angle ($\psi$) as a function of the strength of the  DMI. Solid line corresponds to the numerically calculated minimum based on~Eq.~\eqref{eq:etot-psi}, symbols represent results of the full-scale micromagnetic simulations (open square marks). The simulation data where magnetostatics was reduced to the effective anisotropy is shown with symbols. (d) A structure of the domain wall for two different DMI parameters, see mark in panel~(c). (e) Size of bistability regions $d_0^\text{bis}$ for different angles $\alpha$ of the in-plane easy axis $\vec{e}_2$.
	}
	\label{fig:intro}
\end{figure*}

We consider an infinitely long magnetic thin stripe of thickness $h$ and width $w$. The total magnetic energy of the stripe is
$E = h \int \mathrm{d}S \left[ \mathscr{E}_x + \mathscr{E}_\textsc{a} + \mathscr{E}_\textsc{dm} + \mathscr{E}_\textsc{z} \right],$
where the integration is performed over the sample's area in $xy$ plane ($\vec{\hat{x}}$ axis is along the stripe). The first energy term is the exchange energy density $\mathscr{E}_x = A\sum_{i=x,y,z} (\partial_i \vec{m})^2$ with $A$ being the exchange stiffness, $\vec{m} = \vec{M}/M_\textsc{s}$ being the unit magnetization vector and $M_\textsc{s}$ being the saturation magnetization. The second term is the anisotropy energy density of a biaxial magnet, $\mathscr{E}_\textsc{a} = K_1(1-m_z^2) - K_2 (\vec{m}\cdot \vec{e}_2)^2$, with $K_1 > K_2 > 0$. The easy axis of the in-plane anisotropy $\vec{e}_2$ lies in the stripe's plane at an angle $\alpha$ to the $\vec{\hat{x}}$ direction. The third energy term is the energy density of the DMI\cite{Bogdanov89r,Bogdanov01} $\mathscr{E}_\textsc{dm} = D \left[m_z (\nabla\cdot \vec{m}) - (\vec{m}\cdot \nabla m_z)\right]$. The last energy term is the Zeeman energy density $\mathscr{E}_\textsc{z} = -M_\textsc{s}B m_z$ with $B$ being an external magnetic field intensity. We assume, that the magnetostatic interaction can be reduced to a local anisotropy and results in the renormalization of the first anisotropy constant $K_1 = K_0 - 2\pi M_\textsc{s}^2$ with $K_0$ being the strength of the magnetic anisotropy with an out-of-plane easy axis. 

To describe the structure of the domain wall, we apply the following ansatz\cite{Boulle13a}:
{
\begin{equation}\label{eq:ansatz}
\cos\theta = - p \tanh \xi, \phi = \psi - 90^\circ, \xi = \dfrac{(x-q\ell) \cos \chi + y \sin \chi}{\Delta\ell},
\end{equation}} 
where the magnetization vector is parametrized as $\vec{m} = \left\{ \sin\theta\cos\phi, \sin\theta\sin\phi, \cos\theta \right\}$ in the local spherical reference frame with $\theta$ and $\phi$ being polar and azimuthal angles, respectively, $p = \pm 1$ is the topological charge of the domain wall (kink or anti-kink), $\ell = \sqrt{A/K_1}$ is the magnetic length, $\Delta$ and $q$ are domain wall width and position of its center, respectively, measured in units of $\ell$. The origin is placed in the center of the stripe. The angle $\psi \in (-180^\circ,180^\circ]$ describes the tilt of the magnetization inside the domain wall with respect to $\vec{\hat{y}}$ axis
and the angle $\chi \in (-90^\circ,90^\circ)$ characterizes the mechanical tilt of the plane of the domain wall with respect to $\vec{\hat{y}}$, see Fig.~\ref{fig:intro}(a). In this notation, $\psi = 0$ or $180^\circ$ and $\psi = 90^\circ$ or $-90^\circ$ with $\chi = 0$ corresponds to N\'{e}el and Bloch domain walls, respectively. The plane of the domain wall is perpendicular to the stripe axis when $\chi = 0$. 


The total energy, normalized by $E_0 = 2K_1hw\ell$, reads
\begin{equation}\label{eq:energy-tot-norm}
\begin{split}
\mathcal{E} = \dfrac{E}{E_0} & = \dfrac{1}{\cos\chi} \Bigg\{\frac{1}{\Delta} + \Delta \times \Big[1-k_2 \sin^2(\psi-\alpha)\Big] \\
& + d_0 \sin(\psi-\chi) \Bigg\} - pbq,
\end{split}
\end{equation}
where $k_2 = K_2/K_1$ is the normalized in-plane anisotropy, $d_0 = \pi pD/(2\sqrt{AK_1})$ is the dimensionless parameter characterizing the DMI strength and $b = M_\textsc{s}B/K_1$ is the normalized magnetic field $B$ along $\vec{\hat{z}}$. Note, that the DMI parameter $d_0$ incorporates the topological charge of the domain wall $p$. The static domain wall configuration ($b = 0$, and $q = 0$ without loss of generality) is given by the minimum of the energy~\eqref{eq:energy-tot-norm} with respect to $\Delta$, $\chi$ and $\psi$. The equilibrium domain wall width is $\Delta_0(\psi) = 1/\sqrt{1 - k_2 \sin^2(\psi-\alpha)}$. The relation between values of the angles ${\chi}$ and ${\psi}$ in equilibrium reads
\begin{equation}\label{eq:opt-chi}
2\sin{\chi} = d_0 \Delta_0(\psi)\cos{\psi},
\end{equation}
After the substitution of~\eqref{eq:opt-chi} in~\eqref{eq:energy-tot-norm}, we obtain the expression for the energy as a function of the angle $\psi$, characterizing the orientation of the magnetization in the wall:
\begin{equation}\label{eq:etot-psi}
\mathcal{E}(\psi) = \sqrt{\dfrac{4}{\Delta_0(\psi)^2}-d_0^2 \cos^2\psi} + d_0 \sin\psi.
\end{equation}

\begin{figure*}
	\includegraphics[width=\textwidth]{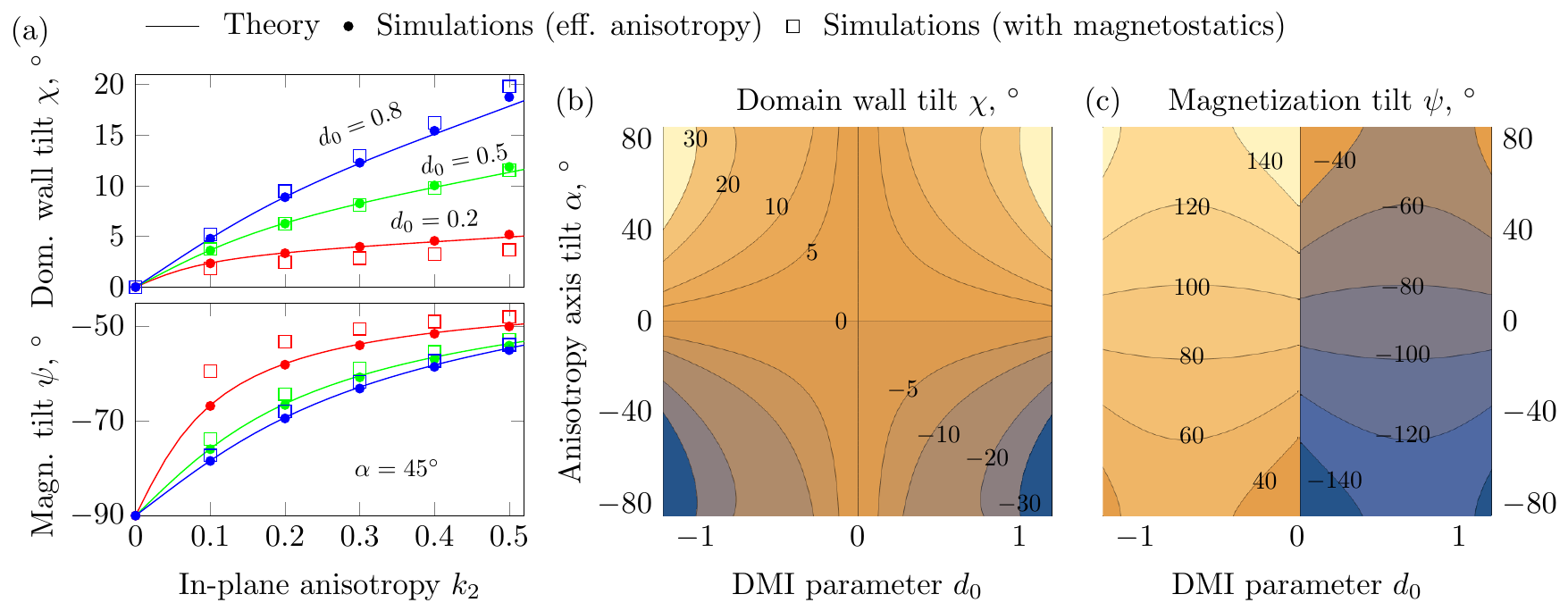}
	\caption{\textbf{The domain wall structure (tilt of magnetization $\psi$ and domain wall tilt $\chi$) as a function of material parameters.} (a) Tilt angles as a function of the strength of the in-plane anisotropy $k_2$ with the easy axis direction $\alpha = 45^\circ$ and different strength of the DMI parameter $d_0$. (b) and (c) Domain wall and magnetization tilt angles for different $d_0$ and $\alpha$. Values of angles are shown with isolines. The normalized anisotropy coefficient $k_2 = 0.30$, topological charge of the domain wall $p = +1$. 
	}
	\label{fig:contours}
\end{figure*}

There are several limiting cases related to the absence of the in-plane anisotropy ($k_2 = 0$) or DMI ($d_0 = 0$). If $k_2 = 0$ and $d_0 = 0$, we obtain a classical case when a magnetic stripe with perpendicular anisotropy can support Bloch domain walls ($\psi = 0,180^\circ$ as a consequence of minimization of magnetostatic energy), with a plane of the domain wall being perpendicular to the stripe axis ($\chi = 0$). For any finite $k_2$ (still when $d_0 = 0$), the magnetization in the domain wall is tilted to its equilibrium value of $\psi = \alpha\pm90^\circ$. This corresponds to the two equivalent minima in the energy~\eqref{eq:etot-psi}, see red line ($d_0 = 0$) in Fig.~\ref{fig:intro}(b). However, the plane of the domain wall remains perpendicular to the stripe axis ($\chi = 0$). This result is expected from the analysis of~\eqref{eq:opt-chi}, indicating that the mechanical rotation of the plane of the domain wall is possible only if the sample possesses a finite DMI. A nonzero DMI results in the symmetry breaking between the opposite magnetization directions, see blue line ($d_0 = 0.1$) in Fig.~\ref{fig:intro}(b). Arrows indicate the shift of the energy minima with the change of the DMI constant. 
For a sufficiently large DMI, the second energy minimum disappears and only one orientation of the domain wall remains stable, see green line ($d_0 = 0.3$) in  Fig.~\ref{fig:intro}(b).

Fig.~\ref{fig:intro}(c) shows the domain wall structure in a broad range of DMI parameters for $k_2 = 0.30$ and $\alpha = 45^\circ$. Solid lines represent the numerically determined minimum of the energy~\eqref{eq:etot-psi} taking into account~\eqref{eq:opt-chi}. The domain walls acquires the \textit{unidirectional tilt} for any finite DMI parameter $d_0$: the energetically preferable state corresponds to $\chi > 0$ only if $\alpha > 0$. The bistability region exists in a vicinity of $d_0 = 0$. 
We made two sets of micromagnetic simulations\footnote{Numerical analysis is performed using energy minimization in OOMMF\cite{OOMMFa,Donahue99,Cortes-Ortuno18} for samples of length 1000~nm, width 200~nm and thickness 1~nm with mesh $2\times2\times 1$~nm with the domain wall placed in the center. The material parameters correspond to Co/Pt ultrathin films with the saturation magnetization $M_\textsc{s} = 1100$ kA/m, exchange stiffness $A = 16$ pJ/m and out-of-plane anisotropy $K_0 = 1.3$ MJ/m$^3$. The in-plane anisotropy coefficient $K_2 = K_0/8$ is chosen if other is not stated. The domain wall structure is extracted from the inner part of the stripe with $|y| < 40$~nm to avoid boundary effects. All data presented in figures are calculated for the case when the easy axis of the in-plane anisotropy is directed at $\alpha = 45^\circ$, if other is not stated. The difference between simulations and analytical theory for is explained by the influence of non-local magnetostatics contribution in the corresponding simulation series and the difference between ansatz~\eqref{eq:ansatz} and real domain wall structure. The finite stripe length also influences  the domain wall structure for very large $\chi$ if magnetostatics is calculated explicitly (full scale micromagnetic simulations).}: symbols show the result, where magnetostatics is reduced to a local anisotropy and open squares represent full-scale simulations. The internal domain wall structure, given by the angle $\psi$, is governed by the direction of the easy axis of the in-plane anisotropy $\vec{e}_2$. In an extended film, the domain wall is always oriented perpendicularly to $\vec{e}_2$ and the DMI energy favors $\psi=\chi$ or $\psi = \chi + 180^\circ$ (magnetization rotates perpendicularly to the domain wall plane). In a stripe of a finite width, the balance between the domain wall tension energy (proportional to its length and increasing with $\chi$) and the DMI energy results in a certain value of $\chi$, which is different from $\psi$. Note that the model~\eqref{eq:ansatz} is applicable for relatively narrow stripes, where the domain wall shape can be approximated by a straight line. For wide stripes the curvilinear distortion of the domain wall shape must be taken into account. The domain wall tilt angle $\chi$ rapidly increases when $d_0$ approaches its critical value. The domain wall structure is shown in Fig.~\ref{fig:intro}(d), where tilt angles $\chi$ and $\psi$ as well as the orientation of the easy axis of the in-plane anisotropy are depicted. The size of the bistability region in terms of the DMI parameter $d_0^\text{bis}$ is shown in Fig.~\ref{fig:intro}(e). Note, that the state $\alpha = 0$ is degenerated with the domain wall tilt $\chi \equiv 0$.

The dependencies of the domain wall ($\chi$) and magnetization ($\psi$) tilt angles on the orientation of the easy axis of the in-plane anisotropy ($\alpha$) is summarized in Fig.~\ref{fig:contours}(b,c). The sign of $\chi$ is given by the direction of the anisotropy axis $\alpha$, while the sign of $\psi$ is opposite to the sign of $d_0$. The domain wall tilt angle monotonically increases with the increase of $d_0$ and $k_2$, while the magnetization tilt angle is mainly determined by the $k_2$ for the case of strong DMI. 


\begin{figure*}
	\includegraphics[width=\textwidth]{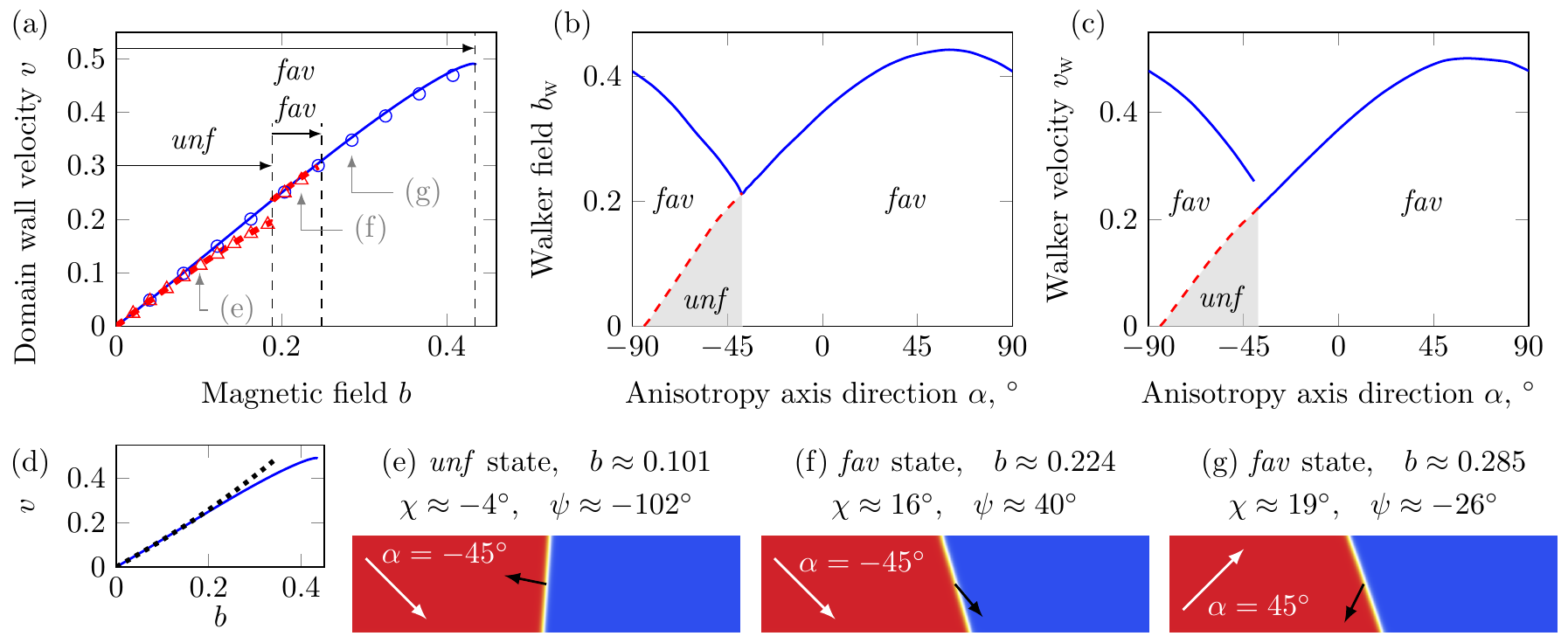}
	\caption{\textbf{Domain wall dynamics.} (a) Comparison of the results of micromagnetic simulations (symbols) and collective variables model (lines). Blue and red colors correspond to direction of the easy axis of the in-plane anisotropy $\alpha = 45^\circ$ and $\alpha = -45^\circ$, respectively. Labels $\psi_\text{fav}$ and $\psi_\text{unf}$ indicate parameters, where favorable and unfavorable magnetization tilt exist, gray arrows indicate simulations, represented in panel (e)--(g). (b) and (c) Dimensionless Walker field and velocity (solid blue lines). Red dashed line shows the field and velocity at which the magnetization tilt $\psi$ changes from unfavorable value to the favorable one (\textit{unf} to \textit{fav}). 
	(d) Comparison of the numerical solution of the collective variable model (solid blue line) with asymptotic (dashed black line). 
	(e)--(g) Structure of a moving domain wall for different orientation of the easy axis of the in-plane anisotropy $\alpha$ and applied field $b$, see also gray arrows in the panel (a). The parameters used for these simulations are $k_2 = 0.30$, $d_0 = 0.63 $, $p = +1$, $\eta = 0.5$.
	}
	\label{fig:walker}
\end{figure*}

In the following, we address the dynamics of domain walls driven by an external magnetic field applied along $\vec{\hat{z}}$. We apply a collective variables approach\cite{Malozemoff79,Slonczewski72}, considering the wall position $q(t)$, the magnetization tilt $\psi(t)$, domain wall tilt $\chi(t)$ and the domain wall width $\Delta(t)$ as time-dependent quantities. The solutions of the corresponding Euler--Lagrange--Rayleigh equations are found numerically, see Supplementary Materials for details and compared with micromagnetic simulations\footnote{We consider stripes of length 2000~nm for simulations of domain wall dynamics and domain wall initial position 500~nm far from the stripe end. Only exchange interaction, anisotropies and DMI are taken into account.}, see Fig.~\ref{fig:walker}(a). The analysis is performed in the fields, which are smaller than the Walker field ($b < b_\textsc{w}$)\footnote{The discussed model is not applicable above Walker fields due to instability of domain wall shape against bending. Micromagnetic simulations show irregular wavy bends of domain boundary during its expansion, see Supplementary Materials.}. In this case, the tilt angles $\psi$ and $\chi$ and the domain wall width quickly relax to their equilibrium values $\psi_\infty$ and $\chi_\infty$, respectively (see also Eqns.~(S-11)--(S-13) in Supplementary Materials). 
The domain wall velocity with equilibrium values of its width and angles  reads
\begin{equation}\label{eq:dynamics}
v = \dfrac{pb}{2\eta \cos\chi_\infty\sqrt{1 - k_2 \sin(\psi_\infty - \alpha)}},
\end{equation}
where the dimensionless velocity $v$ is measured in units of $2\gamma\sqrt{AK_1}/M_\textsc{s}$ with $\gamma$ being gyromagnetic ratio and $\eta$ being Gilbert damping. 
Note, that the maximum of the Walker field and, hence, the largest velocity is reached at the angle $\alpha_0 \approx 60^\circ$ for the given parameters, which does not coincide with a shape anisotropy along the stripe main axis\cite{Yan17,Slastikov19}. Asymptotic analysis shows a good coincidence with numerical solution of equations of motion even for large enough fields and material parameters, see Fig.~\ref{fig:walker}(d) and Supplementary Materials for details.

Upon the motion of the domain wall, its internal structure changes dependent on the direction of the easy axis of the in-plane anisotropy $\alpha$ and on the strength of the applied magnetic field $b$ (Fig.~\ref{fig:walker}). There exists a symmetry break with a favorable magnetization tilt direction (indicated as \textit{fav} states in Fig.~\ref{fig:walker}), in a small angular range about the orientation of $\vec{e}_2$ (along or opposite to it). It results in a higher velocity at a given field. The domain wall with unfavorable tilt angle (indicated as \textit{unf} states in Fig.~\ref{fig:walker}) will switch the internal magnetization in the wall to the $\vec{e}_2$ direction at the beginning of motion. The red dashed line in Fig.~\ref{fig:walker}(b) indicates the smallest field, which is needed for switching of the magnetization angle in the wall. The positive $b$ results in the appearance of ``unf'' state for negative $\alpha$ and vice versa. Sign of $\chi$ coincides with sign of $b$ for fields, close to $b_\textsc{w}$. We note a certain similarity with the texture-induced chirality breaking for moving magnetic domain walls in nanotubes due to non-local magnetostatics\cite{Landeros10,Yan12}.


To summarize, we investigate the internal domain wall structure and its orientation in an out-of-plane mangetized stripe with biaxial (in-plane and out-of-plane) anisotropy and Dzyaloshinskii-Moriya interaction. 
The cooperative effect of the DMI and the additional anisotropy with an in-plane easy axis results in a unidirectional tilt of domain wall in equilibrium and in a symmetry break of a domain wall static state with respect to the stripe axis. The domain wall dynamics in an applied out-of-plane magnetic field exhibits slow and fast motion similarly to vortex domain walls in tubes\cite{Landeros10,Yan12}. We demonstrate that the Walker field but also the associated Walker velocity strongly depend on the orientation of the easy axis of the in-plane anisotropy. There appears an optimal angle of the orientation of the in-plane easy axis to maximize the Walker field and the Walker velocity. This optimal angle does not coincide with the direction of the easy axis of the shape anisotropy. These results are relevant for the optimization of the domain wall dynamics in data storage and logic devices, relying on spintronic and spin-orbitronic concepts.

See Supplementary Material for the details of domain wall dynamics under the action of perpendicular magnetic field.


This work was financed in part via the German Research Foundation (DFG) Grants No. MA5144/9-1, No. MA5144/13-1, MA5144/14-1.
D.D.S. thanks Helmholtz-Zentrum Dresden-Rossendorf e.V., where part of this work was performed, for their kind hospitality and acknowledges the support from the Alexander von Humboldt Foundation (Research Group Linkage Programme).
In part, this work was supported by Taras Shevcheko National University of Kyiv (Project No. 19BF052-01).
Simulations were performed using a high-performance cluster of the Taras Shevchenko National University of Kyiv\cite{unicc}.

\bibliographystyle{aipnum4-1}



\section*{Supplementary material}

# Supplementary materials to "Unidirectional tilt of domain walls in equilibrium in biaxial stripes with Dzyaloshinskii–Moriya interaction"

Oleksandr V. Pylypovskyi,[1,2,a)] Volodymyr P. Kravchuk,[3,4,b)] Oleksii M. Volkov,[1,c)] Jürgen Faßbender,[1,d)] Denis D. Sheka,[2,e)] and Denys Makarov[1,f)]

[1)]*Helmholtz-Zentrum Dresden-Rossendorf e.V., Institute of Ion Beam Physics and Materials Research, 01328 Dresden, Germany*

[2)]*Taras Shevchenko National University of Kyiv, 01601 Kyiv, Ukraine*

[3)]*Institut für Theoretische Festkörperphysik, Karlsruher Institut für Technologie, D-76131 Karlsruhe, Germany*

[4)]*Bogolyubov Institute for Theoretical Physics of National Academy of Sciences of Ukraine, 03143 Kyiv, Ukraine*

(Dated: 10 January 2020)

[a)]Electronic mail: o.pylypovskyi@hzdr.de
[b)]Electronic mail: volodymyr.kravchuk@kit.edu
[c)]Electronic mail: o.volkov@hzdr.de
[d)]Electronic mail: j.fassbender@hzdr.de
[e)]Electronic mail: sheka@knu.ua
[f)]Electronic mail: d.makarov@hzdr.de

## I. ASYMPTOTIC ANALYSIS OF STATIC DOMAIN WALL

Equilibrium values of domain wall and magnetization tilt angles $\chi$ and $\psi$ can be found analytically considering corrections $\Delta\chi$ and $\Delta\psi$ to the case of $d_0 = 0$ and $k_2 > 0$ to be small and by expanding energy (2) to $\Delta\chi^2$ and $\Delta\psi^2$, see main text. In the case of small $d_0$ and $k_2$, the tilt angles read:

$$\chi_{\text{lin}}^{\text{static}} = \pm \frac{2d_0 k_2 \sin\alpha}{\sqrt{1-k_2}(4k_2 + 2d_0\sqrt{1-k_2}\cos\alpha - d_0^2 \cos^2\alpha)},$$
$$\psi_{\text{lin}}^{\text{static}} = \alpha \mp 90° \mp \frac{2d_0\sqrt{1-k_2}\sin\alpha - d_0^2 \sin\alpha\cos\alpha}{4k_2 + 2d_0\sqrt{1-k_2}\cos\alpha - d_0^2 \cos^2\alpha}. \tag{S-1}$$

Note, that both $k_2$ and $d_0$ cannot be simultaneously turned to 0 in (S-1). The numerically calculated energy landscape for Eq. (2) (see main text) is shown in Fig. S-1.

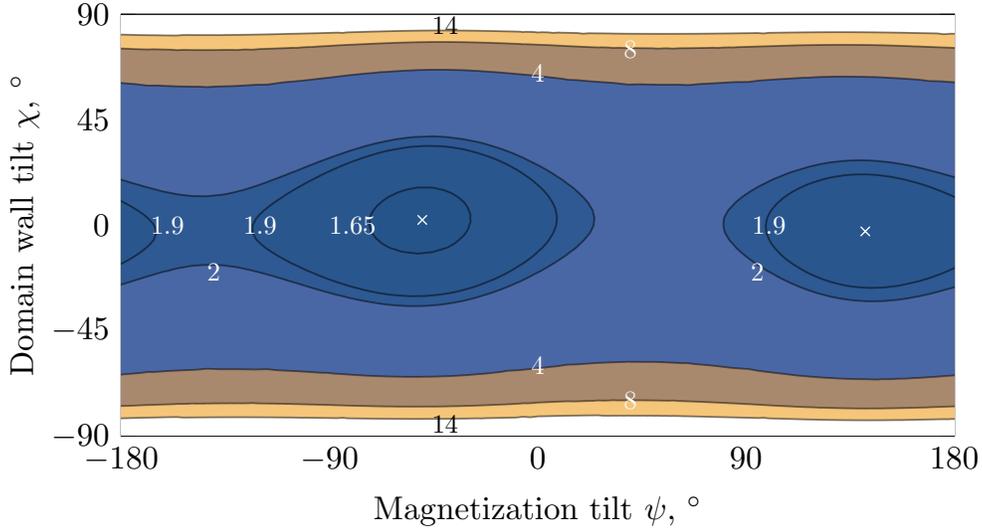

FIG. S-1. **The energy landscape for domain wall.** The landscape by Eq. (2) (see main text) is shown in coordinates of the domain wall tilt $\chi$ and magnetization tilt $\psi$. White crosses mark local energy minima in points with coordinates $(-49.91, 2.20)$ and $(141.5, -2.68)$ with energy values 1.6 and 1.74, respectively. Marks for isolines show values of $\mathscr{E}$. $k_2 = 0.3$, $\alpha = 45°$, $d_0 = 0.1$, $p = +1$.

## II. DOMAIN WALL DYNAMICS

Below the Walker limit $b_{\text{w}}$, the domain wall can move with a steady velocity being straight. Above $b_{\text{w}}$, its motion become irregular, see Figs. S-2 and S-3 for different material parameters.

2

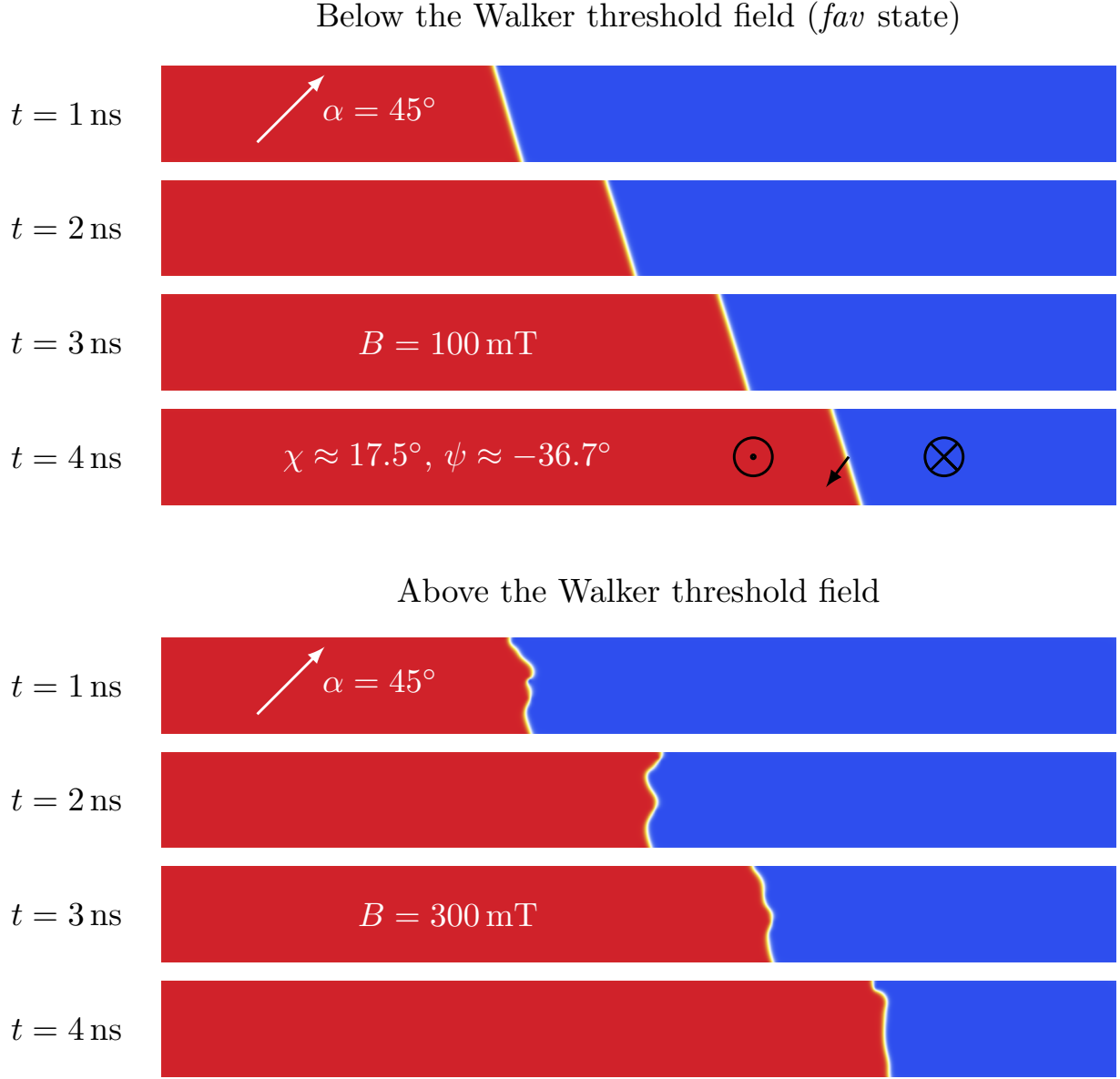

FIG. S-2. **Motion of domain wall below and above Walker threshold in stripe with anisotropy tilt** $\alpha = 45°$. Illustration of domain wall motion, depicted in Fig. 3(a) of the main text (blue line and open symbols). Black arrows indicate magnetization direction. Domain wall moves with a constant velocity without change of structure below the Walker field and randomly bends above it.

We apply the collective variables approach[1,2] to describe the domain wall dynamics in biaxial ferromagnetic stripe in presence of Dzyaloshinskii–Moria interaction, the easy axis of out-of-plane anisotropy perpendicular to the stripe and the easy-axis of in-plane anisotropy tilted by $\alpha$

3

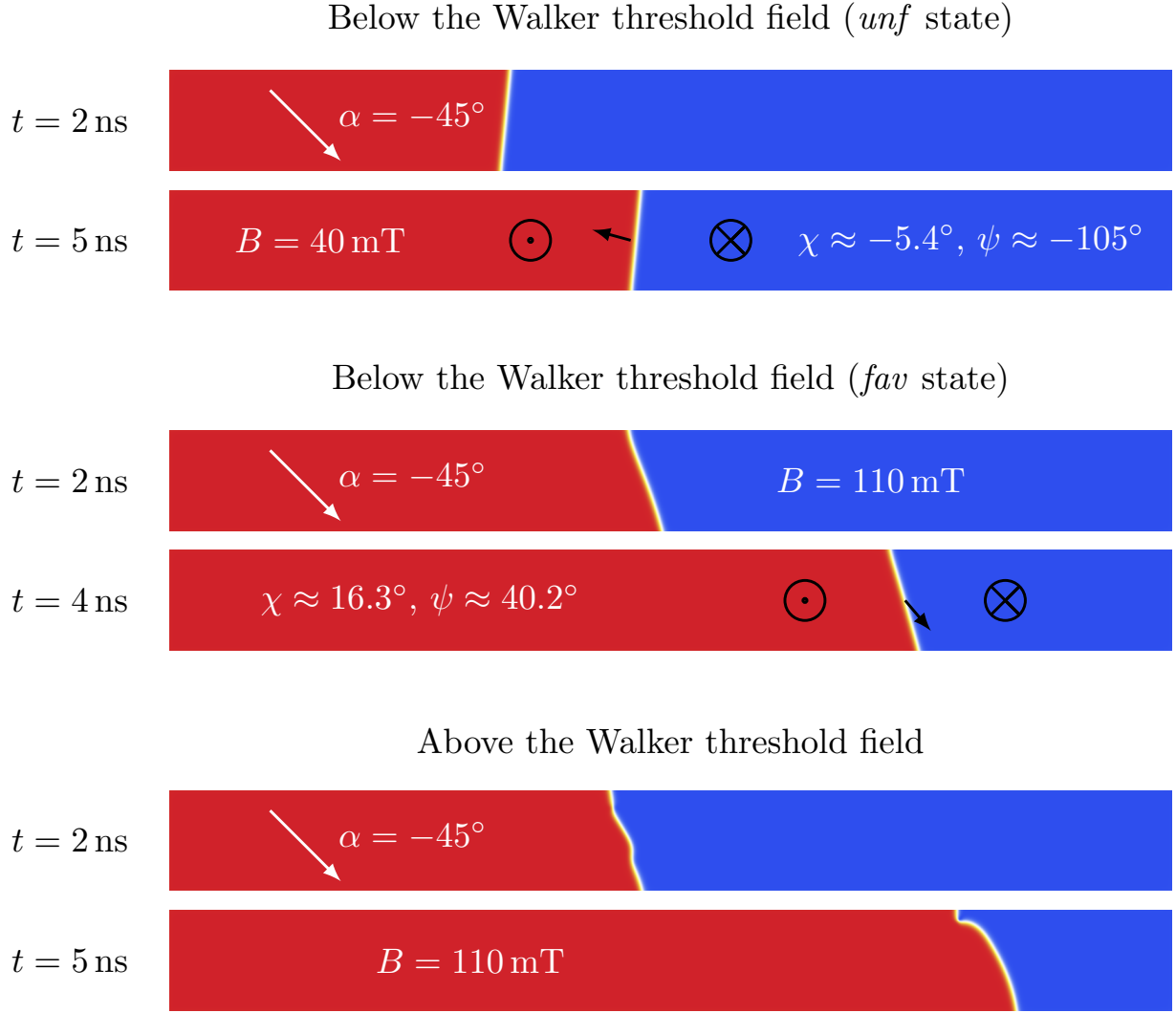

FIG. S-3. **Motion of domain wall below and above Walker threshold in stripe with anisotropy tilt $\alpha = -45°$.** Illustration of domain wall motion, depicted in Fig. 3(a) of the main text (red dashed line and open triangles). Black arrows indicate magnetization direction. Magnetization tilt angle evolves from the equilibrium direction at $b=0$ in *unf* state and rotates to the in-plane easy axis direction in *fav* state. Shape of the domain wall randomly bends above the Walker threshold field.

with respect to the stripe axis $\hat{x}$. The system considered can be described by Lagrangian[3]

$$L = -\frac{M_s h}{\gamma} \int \phi \sin\theta \frac{\partial \theta}{\partial t} dS - E \qquad \text{(S-2)}$$

and dissipative function[4,5]

$$F = \frac{\eta M_s h}{\gamma} \int \left[ \left(\frac{\partial \theta}{\partial t}\right)^2 + \sin^2\theta \left(\frac{\partial \phi}{\partial t}\right)^2 \right] dS, \qquad \text{(S-3)}$$

4

see definitions in main text. The using of ansatz (1) (see main text) gives Euler–Lagrange–Rayleigh equations

$$\frac{\partial \mathscr{L}}{\partial X} - \frac{d}{d\tau}\frac{\partial \mathscr{L}}{\partial \dot{X}} = \frac{\partial \mathscr{F}}{\partial \dot{X}}, \quad X = \{q, \psi, \chi, \Delta\}, \tag{S-4}$$

where $\tau = \omega_0 t$ is the dimensionless time with $\omega_0 = 2K_1\gamma/M_s$ and dot means derivative with respect to $\tau$. The normalized Lagrange and dissipative functions read

$$\mathscr{L} = \frac{L}{E_0} = 2p\psi\dot{q} - \frac{1}{\cos\chi}\left\{\frac{1}{\Delta} + \Delta\left[1 - k_2\sin^2(\psi - \alpha)\right] + d_0\sin(\psi - \chi)\right\} + pbq, \tag{S-5}$$

where $E_0 = 2K_1 hw\ell$, and

$$\mathscr{F} = \frac{F}{\omega_0 E_0} = \frac{\eta\Delta}{\cos\chi}\left[\frac{\cos^2\chi}{\Delta^2}\dot{q}^2 + \dot{\psi}^2 + \frac{1}{12}\left(\pi^2\tan^2\chi + \frac{\tilde{w}^2}{\cos^2\chi\Delta^2}\right)\dot{\chi}^2 + \frac{\pi^2}{12}\frac{\dot{\Delta}^2}{\Delta^2}\right.$$
$$\left. + \frac{\pi^2}{6}\frac{\tan\chi}{\Delta}\dot{\chi}\dot{\Delta}\right] \tag{S-6}$$

with $\tilde{w} = w/\ell$ being the dimensionless stripe width. Then, (S-4) read

$$2\eta\frac{\cos\chi}{\Delta}\dot{q} + 2p\dot{\psi} = pb, \tag{S-7}$$

$$2p\cos\chi\dot{q} - 2\eta\Delta\dot{\psi} = d_0\cos(\psi - \chi) - k_2\Delta\sin 2(\psi - \alpha) \tag{S-8}$$

$$\frac{\eta}{6}\left(\pi^2\tan^2\chi\Delta^2 + \frac{\tilde{w}^2}{\cos^2\chi}\right)\dot{\chi} + \frac{\eta}{6}\pi^2\tan\chi\Delta\dot{\Delta} = d_0\frac{\cos\psi}{\cos\chi}\Delta - \tan\chi - \left[1 - k_2\sin(\psi - \alpha)^2\right]\tan\chi\Delta^2 \tag{S-9}$$

$$\frac{\pi^2\eta}{6}\left(\tan\chi\dot{\chi} + \frac{\dot{\Delta}}{\Delta}\right) = \frac{1}{\Delta^2} + k_2\sin^2(\psi - \alpha) - 1. \tag{S-10}$$

Below the Walker limit, the equilibrium values of domain wall width and tilt as functions of equilibrium magnetization tilt $\psi_\infty$ are

$$\Delta_\infty = \Delta(t \to \infty) = \frac{1}{\sqrt{1 - k_2\sin^2(\psi_\infty - \alpha)}} \tag{S-11}$$

and

$$\sin\chi_\infty \equiv \sin\chi(t \to \infty) = \frac{d_0\cos\psi_\infty}{2\sqrt{1 - k_2\sin(\psi_\infty - \alpha)}}. \tag{S-12}$$

The value of $\psi_\infty$ can be found numerically from the following equation:

$$d_0\cos(\chi_\infty - \psi_\infty)\sqrt{1 - k_2\sin^2(\psi_\infty - \alpha)} - k_2\sin 2(\psi_\infty - \alpha) = \frac{b}{\eta}. \tag{S-13}$$

Comparison of numerically found $\chi_\infty$ and $\psi_\infty$ from Eqns. (S-12) and (S-13) with simulations for different material parameters is shown in Fig. S-4.

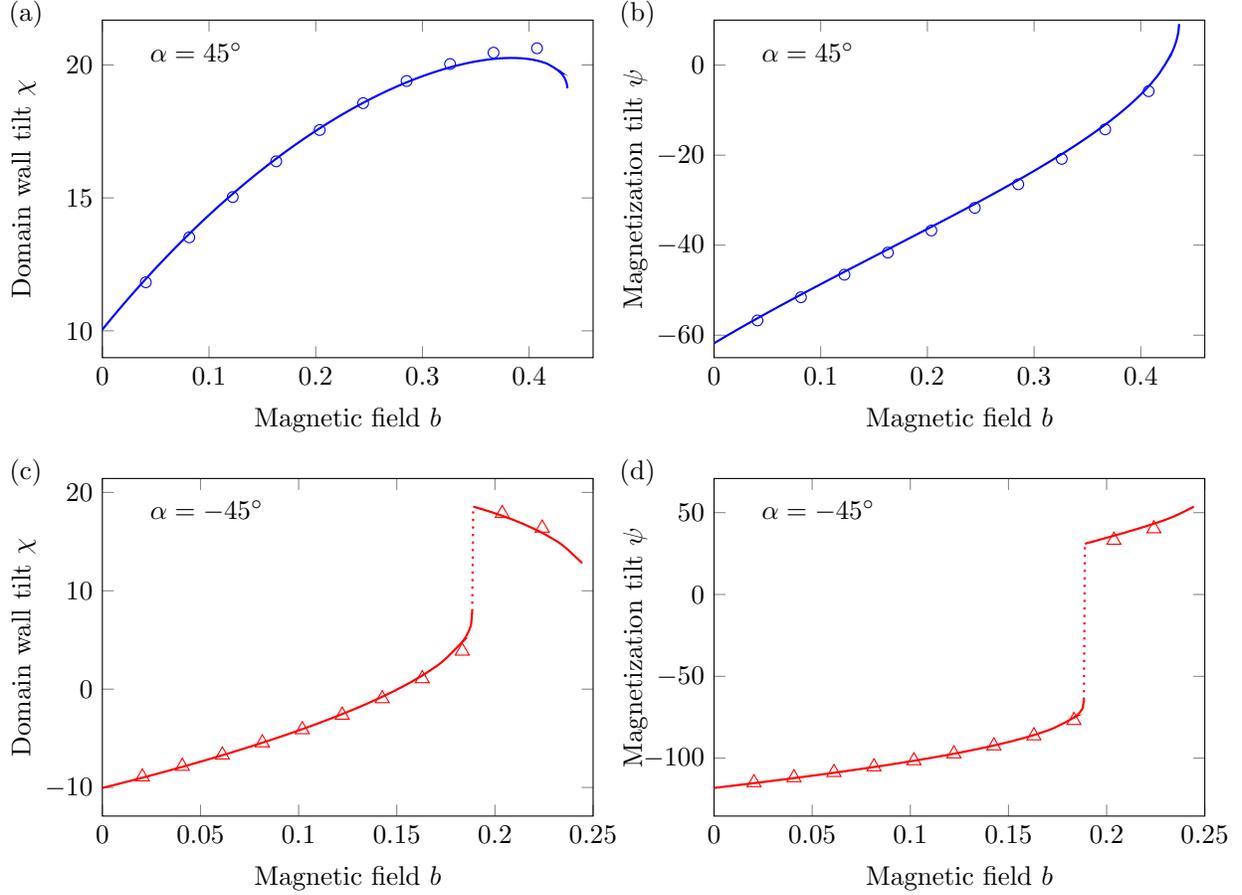

FIG. S-4. **Domain wall and magnetization tilt angles of a moving domain wall.** (a) and (b) Tilt angles for $\alpha = 45°$, see Fig. 3(a) of the main text (blue lines and open symbols). (c) and (d) Tilt angles for $\alpha = -45°$, see Fig. 3(a) of the main text (red dashed lines and open triangles).

Considering case of small deviations of values of $\chi_{\text{static}}^{\text{lin}}$ and $\psi_{\text{static}}^{\text{lin}}$ during the domain wall motion, one can found the following asymptotic expressions for the stable branch:

$$\begin{aligned}\chi_{\infty \text{ lin.}} &= \chi_{\text{st. lin.}} + \frac{bd_0 \cos\alpha}{\eta\left[4k_2\sqrt{1-k_2} + d_0(2 - 2k_2 - d_0\sqrt{1-k_2}\cos\alpha)\cos\alpha\right]}, \\ \psi_{\infty \text{ lin.}} &= \psi_{\text{st. lin.}} + \frac{2b}{\eta\left[4k_2 + d_0(2\sqrt{1-k_2} - d_0\cos\alpha)\cos\alpha\right]},\end{aligned} \quad \text{(S-14)}$$

where values for the stable branch of $\chi$ and $\psi$ (+ and − sign, respectively) are used. Comparison of asymptotics (S-14) with numerical solution of (S-7)–(S-10) for different material parameters is shown in Fig. S-5.

6

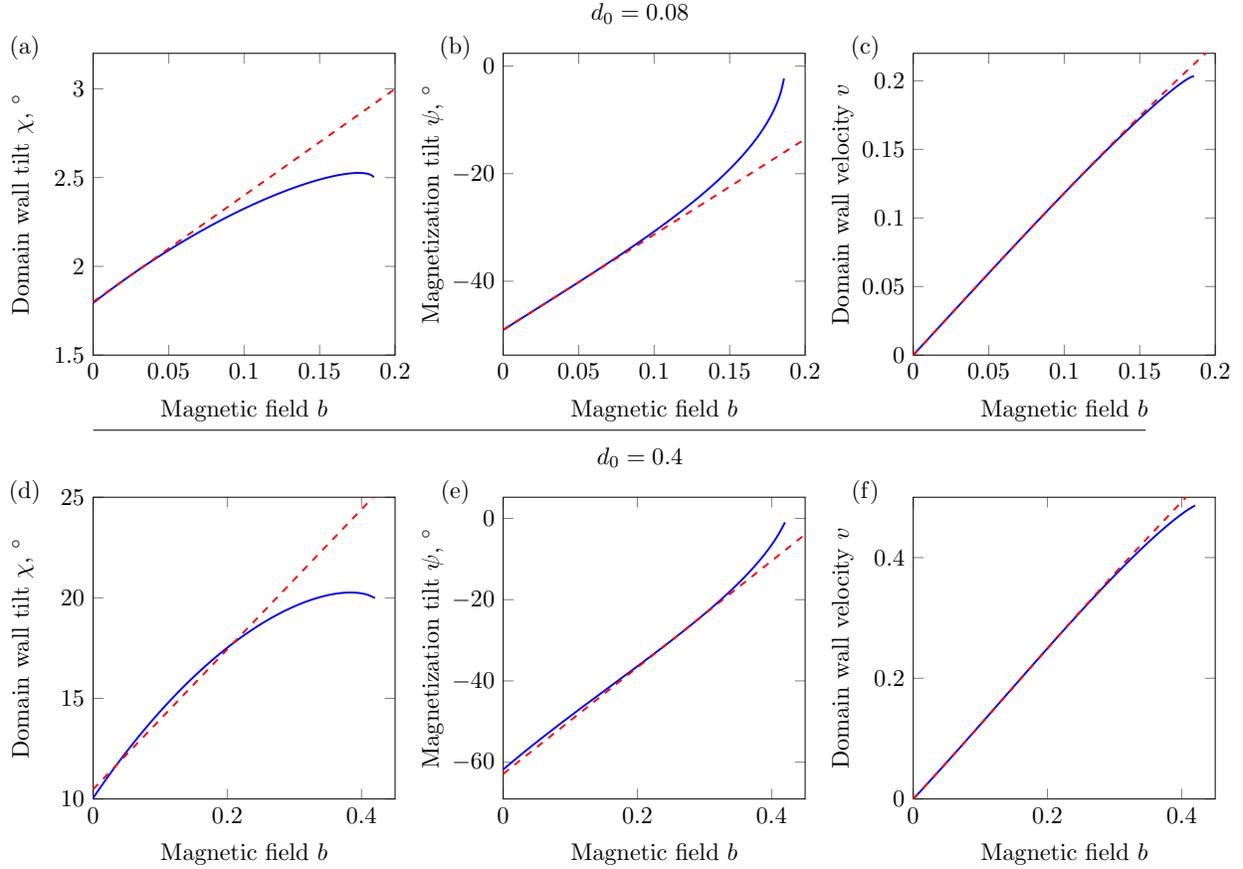

FIG. S-5. **Numerical solution of** (S-7)–(S-10) **(solid blue lines) and asymptotics** (S-14) **(see also main text, Fig. 3(d)) for different magnetic fields below the Walker limit.** (a),(d) Domain wall tilt $\chi$. (b),(e) Magnetization tilt $\psi$. (c),(f) Domain wall velocity. The parameters used are $k_2 = 0.3$, $\alpha = 45°$, $p = +1$, $\eta = 0.5$.

tions on Magnetics **40**, 3443–3449 (2004).



\end{document}